# Role of the dielectric constant of ferroelectric ceramic in enhancing the ionic conductivity of a polymer electrolyte composite

Pramod Kumar Singh[a*] and Amreesh Chandra[b]

(a) Department of Physics, Banaras Hindu University, Varanasi-221005, India

   E mail: singhpk71@rediffmail.com

(b) School of Materials Science and Technology, Institute of Technology, Banaras Hindu University, Varanasi-221005, India

   E mail : rishu_chandra@yahoo.com

Abstract

The dispersal of high dielectric constant ferroelectric ceramic material $Ba_{o.7}Sr_{o.3}TiO_3$ ($T_c \approx 30^0C$) and $Ba_{0.88}Sr_{0.12}TiO_3$ ($T_c \approx 90^0C$) in an ion conducting polymer electrolyte (PEO: $NH_4I$) is reported to result in an increase in the room temperature ionic conductivity by two orders of magnitude. The conductivity enhancement "*peaks*" as we approach the dielectric phase transition of the dispersed ferroelectric material where the ε changes from ~ 2000 to 4000. This establishes the role of dielectric constant of the dispersoid in enhancing the ionic conductivity of the polymeric composites.

Ion conducting polymers, obtained by polar polymer-salt complexation are becoming increasingly important because of their application in solid state polymer batteries and fuel cells. One class of polymer electrolytes which are widely studied is the "polymer-salt complex" in which a dissociable salt is complexed with a polar polymer (like polyethylene oxide PEO;polypropylene oxide PPO etc ). Many good lithium ion conducting polymer electrolytes have been developed [1-4] by complexing polar polymers (say PEO) with lithium salts like lithium perchlorate, lithium triflate, lithium tetrafluoroborate ($LiBF_4$) etc. Apart from $Li^+$-ion conducting polymers, good proton *($H^+$)* ion conducting polymer electrolytes have also been obtained by complexing ammonium salts [5-7] with polar polymers. To enhance the ionic conductivity, various approaches which have been pursued are to: (a) change the type of complexing polymer (b) use polymers with different chain lengths (c) use different complexing salts (d) modify the degree of crystallinity by the use of plasticizers and copolymerisation (e) form "composites" by the dispersal of neutral/chemically inert fillers. Forming "composites" by the dispersal of ceramic fillers is a very promising approach as it can enhance the conductivity as well as provide mechanical strength to the polymeric film. The mechanism of conductivity enhancement has been extensively studied [8-10] for crystalline -crystalline composites. The two important mechanisms suggested for the conductivity enhancement of these composites are: (a) Adsorption-Desorption model of Maier[8] who proposed that the ionic charge carrier concentration at the interface increases in the composites and (b) Percolation model due to Bunde et al[10] who introduced the idea of the presence of a high conductivity path at the interface which develops connectivity or "percolation" at a certain composition threshold of the dispersoid. The situation in the case of the polymer composites is more complicated because the dispersal of a "filler" changes the polymeric structure/morphology. Some of the important additional factors to be considered for the polymeric composites are:

(i) change in glass transition temperature ($T_g$) and melting temperature of the polymer ($T_m$)

(ii) change in the degree of crystallinity

(iii) size of the dispersoid particle because the interface contact area changes inversely as the radius of the particle (1/r) since the surface to volume ratio is approximately proportional to (l/r).

(iv) nature of the dispersoid which may be acidic or alkaline or amphoteric.

We show in this paper that the dielectric constant of the dispersoid is also a controlling factor. To fulfil our motivation and to establish unambiguously the role of dielectric constant of the dispersoid in modifying the ionic conductivity of the polymer electrolyte, we have designed the composite for our study such that it avoids the confusion arising out of the factors listed above as explained later in this paper.

The dielectric constant of the dispersoid is expected to change the ion transport in the polymer electrolyte since its presence modifies the local electric field distribution inside the polymeric composite ionic matrix. In the presence of dispersed "ferroelectric" particles of dielectric constant higher than that of the basic polymer electrolyte ionic matrix, the electric field lines get modified such that the field would be higher "near the interface" of the polymer due to the high dielectric constant dispersoid (i.e., ferroelectric). This can have either of the following two effects:

(a) the "effective" conductivity due to the movement of the mobile ions, near the interface may increase, as the effective local field for the same applied potential is higher, or

(b) the dissociation of the complexed salt[11] in the polymer near the interface may increase resulting in an increase in the number of free mobile charge carriers 'n' (and hence enhancement of the net conductivity) since $n = n_o \exp(-U/2\varepsilon kT)$; where U is the dissociation energy.

In general, the high dielectric constant of the dispersoid may enhance n or effective µ or both resulting in an increase in the overall conductivity.

For experimentally verifying the above statements, a possible experiment could be to choose "different dispersoid materials with widely different dielectric constants". An argument against this is that the resulting effect in the conductivity enhancement would not only be due to the dielectric constant but also will be the combined effect of the factors (i) to (iv) discussed above in an earlier paragraph. All the factors (including dielectric constant) are not mutually exclusive and hence it is experimentally difficult to "unambiguously establish" the role of each separately. This constraint forces us not to change the type of dispersoid. Yet our requirement is to create a condition of measurement such that the dielectric constant "changes significantly". To obtain these conditions, we chose a "ferroelectric material" as dispersoid with "dielectric phase transition" in a temperature range over which the changes in the conductivity of polymer electrolyte can be easily measured. The polymer electrolyte chosen by us is PEO:$NH_4I$ (80:20 wt% )[7] in which a modified $BaTiO_3$ ferroelectric ceramic ($Ba_{1-x}Sr_xTiO_3$) is dispersed. It has been earlier found[12] that by changing x, the dielectric phase transition temperature changes and can be so adjusted that it falls in the desired temperature range in which the conductivity can be easily measured. The choice of 'x' (and corresponding transition temperature) of the dispersoid is "crucial" to our experiment since PEO has $T_m$~ 60°C above which the film becomes amorphous and the conductivity rises suddenly. We have chosen two values of 'x', viz., x=0.30 with $T_c$ ~ $30^0C$ which is lower than the $T_m$ of PEO and x=0.12 with $T_c$ ~ 90°C which is above the $T_m$ of PEO.

The $Sr^{2+}$ modified $BaTiO_3$ ferroelectric ceramic, abbreviated as BST, was prepared using a novel semi-chemical route[12]. Fine powders of BST is obtained by crushing the sintered pellets in agate pistle-mortar and then sieving it by using a 400 mesh sieve. Before dispersing this in polymer, a highly viscous solution of the complexed polymer electrolyte is first obtained. The PEO was dissolved in methanol to which 20 wt % of NH$_4$I was added and stirred at 40°C by a magnetic stirrer for 4 -5 hours. Then, the fine powder of BST(3wt%) was added and again stirred thoroughly till the solution became highly viscous. This viscous solution was poured in a polypropylene petri dish and solution casted films were obtained after thoroughly drying in air followed by vacuum drying. Our earlier studies on the composition dependence of this composite has shown[11] that the composite with 3 wt% of BST has highest conductivity and so this composite was chosen for the present work. The bulk conductivity was evaluated by complex impedance spectroscopy[13,14] technique using Hioki Hi Tester model 3520 in the frequency range 40 Hz-100 KHz. The dielectric constant was measured using Hioki Hi Tester 3532.

The values of the dielectric constant of BST at different temperatures for x=0.30 and x=0.12 are shown as "inset" of Figures 1 and 2. The dielectric constant for x=0.30 sample is nearly 2000 at the low temperature end i.e., < 20°C. It suddenly peaks to a value of 4000 at ~ 30°C and then decreases again to 2000 or lower. For the case of BST with x=0.12, we observe a similar change in the dielectric constant at its transition temperature T$_c$ ~ 90°C.

The σ Vs. 1/T plot for (PEO:NH$_4$I)+ 3 wt% Ba$_{1-X}$ Sr$_x$TiO$_3$ is shown in Figure 1 for the polymeric composite with dispersoid having x=0.30 with T$_c$ =30°C. We see that the conductivity increases as we approach T$_c$≈30°C where the dielectric constant peaks. Above 30°C, the conductivity again reaches its normal value as the dielectric constant becomes nearly equal to its value before the transition temperature T$_c$. The conductivity then shoots up sharply at ~ 50-55°C as we approach the T$_m$ value of the polymer PEO. This increase in the conductivity at T$_m$ is a general observation and has been assigned to the high amorphicity beyond T$_m$.

For the polymer samples containing BST ferroelectric dispersoid with x=0.12; $T_c \sim 90°C$, the change in the conductivity in the 30° - 40°C range is not seen (see Fig. 2) since in this temperature range the values of the dielectric constant does not change much. However, we do see a small hump in the values of σ as we pass through the dielectric phase transition of the BST ceramic (with $T_c \sim 90°C$). This change in the conductivity is not as clearly seen as that in Fig. 1 because beyond $T_m$ the conductivity of the base polymer electrolyte itself is large and the changes in σ introduced by the dispersoid BST are masked.

In brief, we have conclusively shown that in polymer electrolyte- ceramic composites, the high dielectric constant of the dispersed ferroelectric ceramics has a definite role in the enhancement of the conductivity of the composite, possibly as a result of the high effective electric field at the dispersed-polymer interfaces.

Acknowledgement

We thank Professor S. Chandra for valuable guidance and advice. One of us (PKS) thanks C.S.I.R. (Govt. of India) for the award of Senior Research Fellowship under Emeritus Scientist project of Professor S. Chandra.

Figure captions

FIG. 1. Temperature dependence of conductivity for (PEO:NH$_4$I)+ 3 wt% Ba$_{0.70}$ Sr$_{0.30}$TiO$_3$. The inset shows the temperature dependence of the dielectric constant of BST at a frequency of 1 KHz.

FIG. 2. Temperature dependence of conductivity for (PEO:NH$_4$I)+ 3 wt% Ba$_{0.88}$ Sr$_{0.12}$TiO$_3$ sample. The inset shows the values of the dielectric constant of the dispersed BST material at a frequency of 1 KHz.

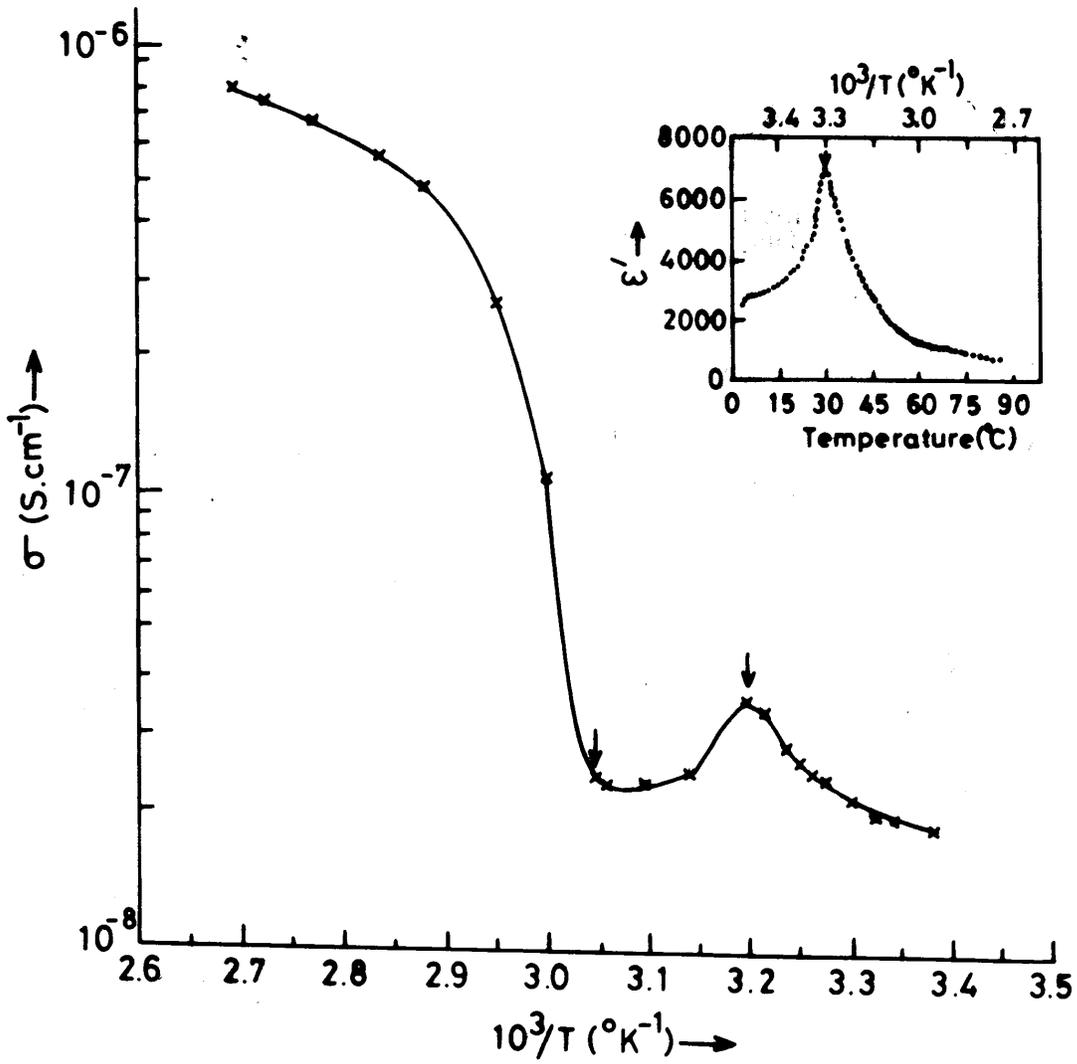

Fig.1

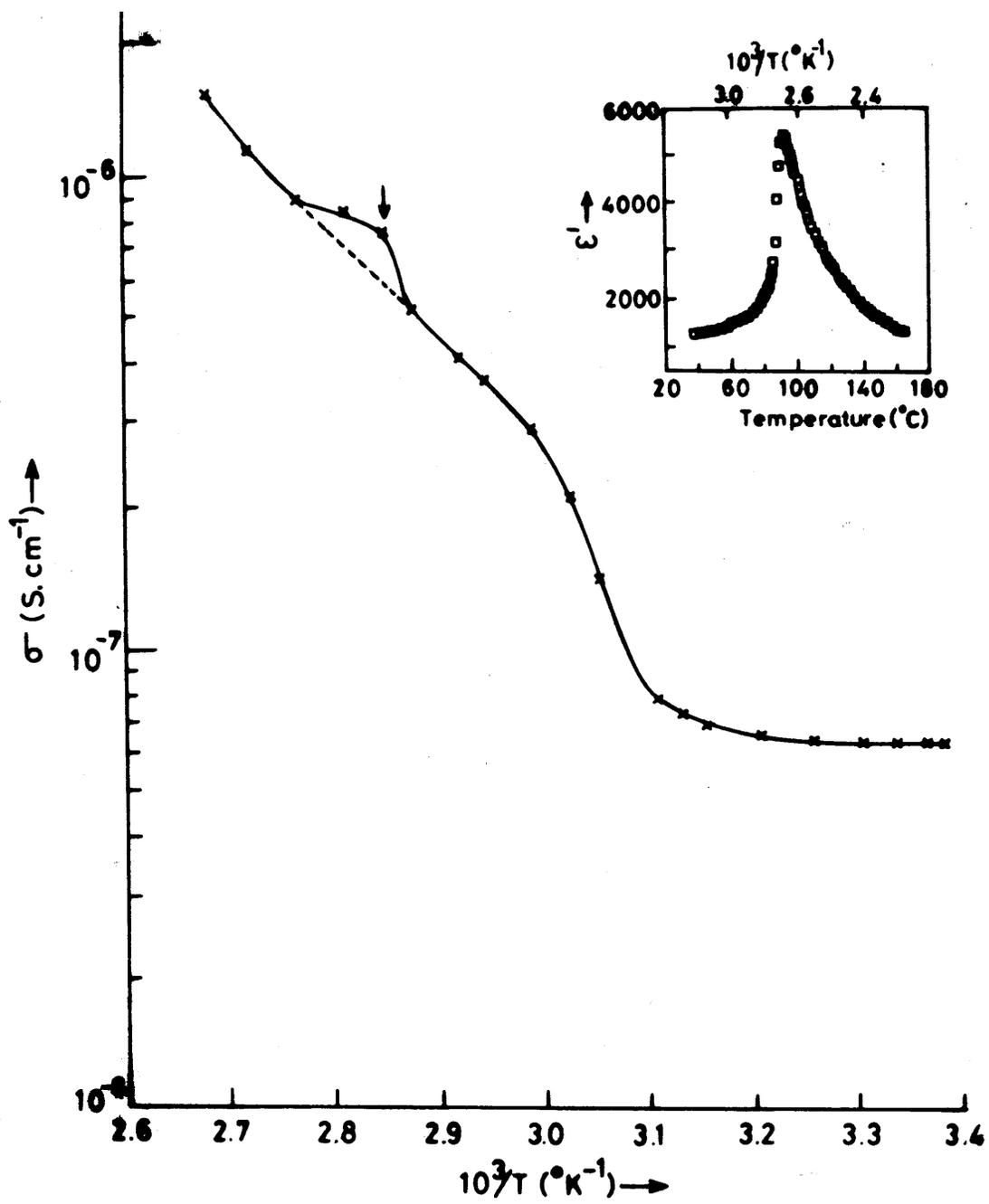

Fig. 2